\documentclass[letterpaper,english,reprint, aps,superscriptaddress]{revtex4-1}
\usepackage[T1]{fontenc}
\usepackage[latin9]{inputenc}
\usepackage{amsmath}
\usepackage{amssymb}
\usepackage{graphicx}
\usepackage{soul}
\usepackage{colortbl}
\usepackage[colorlinks=true,citecolor=blue,linkcolor=blue]{hyperref}

\makeatletter

\pdfpageheight\paperheight
\pdfpagewidth\paperwidth

\usepackage{mathdots}

\makeatother

\usepackage{babel}
\begin{document}

\title{Low-depth circuit ansatz for preparing\\
correlated fermionic states on a quantum computer}

\author{Pierre-Luc Dallaire-Demers}
\affiliation{Xanadu, 372 Richmond St W, Toronto, M5V 2L7, Canada}
\author{Jonathan Romero}
\affiliation{Department of Chemistry and Chemical Biology, Harvard University,
Cambridge MA, 02138}
\author{Libor Veis}
\affiliation{J. Heyrovsk\'{y} Institute of Physical Chemistry, ASCR, 18223 Prague,
Czech Republic}
\author{Sukin Sim}
\affiliation{Department of Chemistry and Chemical Biology, Harvard University,
Cambridge MA, 02138}
\author{Al\'{a}n Aspuru-Guzik}
\affiliation{Department of Chemistry and Chemical Biology, Harvard University,
Cambridge MA, 02138}
\affiliation{Senior Fellow, Canadian Institute for Advanced Research, Toronto, Ontario M5G 1Z8, Canada}

\date{\today}
\begin{abstract}
Quantum simulations are bound to be one of the main applications of
near-term quantum computers. Quantum chemistry and condensed matter
physics are expected to benefit from these technological
developments. Several quantum simulation methods are known to prepare a state
on a quantum computer and measure the desired observables. The most
resource economic procedure is the variational quantum eigensolver (VQE),
which has traditionally employed unitary coupled cluster as the ansatz to approximate ground states of many-body fermionic Hamiltonians. A significant caveat of the method is that the
initial state of the procedure is a single reference product state
with no entanglement extracted from a classical Hartree-Fock calculation.
In this work, we propose to improve the method by initializing the
algorithm with a more general fermionic Gaussian state, an idea borrowed
from the field of nuclear physics. We show how this Gaussian reference
state can be prepared with a linear-depth circuit of quantum matchgates.
By augmenting the set of available gates with nearest-neighbor phase
coupling, we generate a low-depth circuit ansatz that can accurately
prepare the ground state of correlated fermionic systems. This extends
the range of applicability of the VQE to
systems with strong pairing correlations such as superconductors,
atomic nuclei, and topological materials.
\end{abstract}
\maketitle

\section{\label{sec:Intro}Introduction}

The macroscopic properties of matter emerge from its microscopic quantum
constituents whose massive components are mostly fermions. Understanding
and modeling the behavior of a large number of interacting fermions
is a central and fundamental problem in Physics and Chemistry which
requires a large investment in computational resources as the memory
required to represent a many-body state scales exponentially with
the number of particles. Therefore, a computer operating on quantum
mechanical principles have the potential to revolutionize the simulation
of quantum systems \citep{Feynman82,Abrams97}. Such a machine would
improve our ability to design new molecules such as drugs and catalysts \citep{Reiher17}, 
build new superconducting \citep{Bauer15,Wecker15,DDW16}
and topological materials and improve our understanding of nuclear
matter. Algorithm leveraging the advantages of quantum computers for
quantum simulations have steadily been developed in the past two
decades \citep{Ortiz01,Aspuru05,Kassal08,Whitfield11,Kassal11,Veis12,LasHeras13,Peruzzo14,Barends15,Bauer15,LasHeras15,OMalley15,Poulin15,Wecker15,Babbush16,Zhu17}
as quantum processors are scaling in size \citep{Monz16,Linke17,Neill17}.
Variational quantum eigensolvers (VQE) have recently appeared as a
promising class of quantum algorithms designed to prepare states for
quantum simulations \citep{Wecker15b,OMalley15,McClean16}. However, near-term
devices will suffer from limited coherence as a consequence of noise
and finite experimental precision \citep{Temme16,Benjamin17}. This
incentives the search for low-depth circuits for quantum simulations
and state preparation \citep{Kandala17,Babbush17}.

In this paper, we present a new type of low-depth VQE ansatz motivated
by the Bogoliubov coupled cluster theory \citep{Stolarczyk10,Rolik14,Signoracci15}. 
Our approach can be used to prepare the ground state of correlated fermions
with pairing interactions by systematically appending variational
cycles composed of linear-depth blocks of 2-qubit gates. In section
\ref{sec:GVQE}, we first review the formulation of the strong correlation
problem for fermions in the context of second quantization. We then
present the unitary version of Bogoliubov coupled cluster theory and
review how the generalized Hartree-Fock (GHF) reference state can
be computed as a fermionic Gaussian state. Using the theory of matchgates,
we show how pure fermionic Gaussian states can be exactly prepared
on a quantum computer using a linear-depth circuit. Finally, we introduce the low-depth circuit ansatz (LDCA), consisting of the previous matchgate circuit plus additional nearest-neighbor phase coupling. We numerically benchmark the LDCA in section \ref{sec:Numerics} for the prototypical examples of the Fermi-Hubbard model in condensed matter and the automerization reaction of cyclobutadiene in quantum chemistry, showing its potential to describe the exact ground state of strongly correlated systems.

\section{Generalized variational quantum eigensolver\label{sec:GVQE}}

In this section, we review and extend the theoretical foundations
of VQE. Specifically, in subsection \ref{subsec:ProblemDefinition},
we review the definition of finding the ground state of fermionic
Hamiltonians as found in quantum chemistry, condensed matter, and nuclear
physics. In subsection \ref{subsec:BUCC} we introduce the Bogoliubov
unitary coupled cluster (BUCC) theory as a variational ansatz to the
ground state problem. In subsection \ref{subsec:GHFT} we review the
formalism of the GHF theory as this is the starting point of the BUCC
optimization method as well as the new method presented in the following
subsection. In subsection \ref{subsec:QuantumAlgorithm} we show how
a GHF state can be prepared on a quantum processor using matchgates
and introduce a LDCA which can be used to prepare the ground state
of fermionic Hamiltonian with surprisingly high accuracy. Finally, in subsection \ref{subsec:GradientEvaluation}, we outline an implementation to compute the analytical gradient of the LDCA using quantum resources.

\subsection{Formulation of the problem\label{subsec:ProblemDefinition}}

Many systems in quantum chemistry \citep{Attila96}, condensed matter
\citep{Rickayzen91,Senechal04,Leggett06}, and nuclear structure physics
\citep{Ring80,Bender03} can be modeled by an ensemble of interacting
fermions (electrons, nucleons) described by a second quantized Hamiltonian
of the form
\begin{equation}
\begin{array}{rcl}
H & = & \sum_{pq}\left(t_{pq}a_{p}^{\dagger}a_{q}+\Delta_{pq}a_{p}^{\dagger}a_{q}^{\dagger}+\Delta_{pq}^{\ast}a_{q}a_{p}\right)\\
\\
 &  & +\sum_{pqrs}v_{pqrs}a_{p}^{\dagger}a_{q}^{\dagger}a_{s}a_{r}\\
\\
 &  & +\sum_{pqrstu}w_{pqrstu}a_{p}^{\dagger}a_{q}^{\dagger}a_{r}^{\dagger}a_{u}a_{t}a_{s}.
\end{array}\label{eq:Hamiltonian}
\end{equation}

In general, the $p$,$q$,$\ldots$,$u$ indices run over all relevant
quantum numbers (e.g. position, momentum, band number, spin, angular
momentum, isospin, etc) which define $M$ fermionic modes. The fermionic
mode operators follow canonical anti-commutation relations $\left\{ a_{k},a_{l}^{\dagger}\right\} =\delta_{kl}$
and $\left\{ a_{k},a_{l}\right\} =\left\{ a_{k}^{\dagger},a_{l}^{\dagger}\right\} =0$.
The kinetic energy terms $t_{pq}$ and the interaction $v_{pqrs}$
are ubiquitous in most theories, while pairing terms $\Delta_{pq}$
often appear in the context of mean-field superconductivity, and the
three-body interaction term $w_{pqrstu}$ can be phenomenologically
introduced in nuclear physics \citep{Loiseau67}.

As a prerequisite to calculating various observable quantities, we
are interested in finding the ground state $\rho_{0}=\left|\Psi_{0}\right\rangle \left\langle \Psi_{0}\right|$
of the Hamiltonian \eqref{eq:Hamiltonian} such that the energy $E$
is minimized over the set of all possible states $\rho$ in a given
Hilbert space:
\begin{equation}
\begin{array}{rcl}
E_{0} & \equiv & E\left(\rho_{0}\right)\\
\\
 & = & \textrm{min}_{\rho}E\left(\rho\right)\\
\\
 & = & \textrm{min}_{\rho}\textrm{tr}\left(H\rho\right).
\end{array}\label{eq:GroundStateProblem}
\end{equation}
When this minimization cannot be done either analytically or with
numerically exact methods, we have to resort to approximate methods
such as variational ansatzes. One such ansatz, the BUCC method, is defined
in the next subsection. 

\subsection{Bogoliubov unitary coupled cluster theory\label{subsec:BUCC}}

Coupled cluster methods are used in ab initio quantum chemistry calculations to describe correlated many-body states with a better accuracy than the Hartree-Fock method. Bogoliubov- and quasiparticle-based coupled
cluster methods extends the range of applicability of those methods
to systems with mean-field paired states \citep{Stolarczyk10,Rolik14,Signoracci15}.
Anticipating the implementation on quantum computers, we present the
formalism for the unitary version of the Bogoliubov coupled cluster
theory. We first review the Bogoliubov transformation and the parametrization
of the ansatz.

The most general linear transformation acting on fermionic creation
and annihilation operators that preserves the canonical anti-commutation
relation is the Bogoliubov transformation. In this transformation,
the quasiparticle operators $\left(\beta_{p'}^{\dagger};\beta_{p'}\right)$
are related to the single-particle operators $\left(a_{p}^{\dagger};a_{p}\right)$
by a unitary matrix 
\begin{equation}
\begin{array}{rcl}
\beta_{p'}^{\dagger} & = & \sum_{p}\left(U_{pp'}a_{p}^{\dagger}+V_{pp'}a_{p}\right)\\
\\
\beta_{p'} & = & \sum_{p}\left(U_{pp'}^{\ast}a_{p}+V_{pp'}^{\ast}a_{p}^{\dagger}\right).
\end{array}\label{eq:BogoliubovTransformation}
\end{equation}
This transformation preserves the canonical anti-commutation relation
such that $\left\{ \beta_{k},\beta_{l}^{\dagger}\right\} =\delta_{kl}$
and $\left\{ \beta_{k},\beta_{l}\right\} =\left\{ \beta_{k}^{\dagger},\beta_{l}^{\dagger}\right\} =0$.
By introducing the vector notation $\vec{a}^{\top}=\left(a_{1},\ldots,a_{M},a_{1}^{\dagger},\ldots,a_{M}^{\dagger}\right)$
and $\vec{\beta}^{\top}=\left(\beta_{1},\ldots,\beta_{M},\beta_{1}^{\dagger},\ldots,\beta_{M}^{\dagger}\right)$,
it is easy to express \eqref{eq:BogoliubovTransformation} in matrix
notation as $\vec{\beta}=\mathcal{U}\vec{a}$ where the Bogoliubov
transformation is unitary $\mathcal{U}^{-1}=\mathcal{U}^{\dagger}$
and its matrix is defined as
\begin{equation}
\mathcal{U}=\left(\begin{array}{cc}
\mathbf{U}^{\ast} & \mathbf{V}^{\ast}\\
\mathbf{V} & \mathbf{U}
\end{array}\right).\label{eq:BogoliubovMatrix}
\end{equation}
The ground state of a quadratic Hamiltonian (all $v_{pqrs}=0$ and
$w_{pqrstu}=0$) is a product state
\begin{equation}
\left|\Phi_{0}\right\rangle =C\prod_{k=1}^{M}\beta_{k}\left|\textrm{vac}\right\rangle ,\label{eq:HFBGroundState}
\end{equation}
where $\left|\textrm{vac}\right\rangle $ is the Fock vacuum and $C$
is a normalization factor. If the ground state is not degenerate,
\eqref{eq:HFBGroundState} acts as a quasiparticle vacuum $\beta_{j}\left|\Phi_{0}\right\rangle =0$.

We can define the quasiparticle cluster operator $\mathcal{T}=\mathcal{T}_{1}+\mathcal{T}_{2}+\mathcal{T}_{3}+\ldots$
where
\begin{equation}
\begin{array}{rcl}
\mathcal{T}_{1} & = & \sum_{k_{1}k_{2}}\theta_{k_{1}k_{2}}\beta_{k_{1}}^{\dagger}\beta_{k_{2}}^{\dagger}\\
\\
\mathcal{T}_{2} & = & \sum_{k_{1}k_{2}k_{3}k_{4}}\theta_{k_{1}k_{2}k_{3}k_{4}}\beta_{k_{1}}^{\dagger}\beta_{k_{2}}^{\dagger}\beta_{k_{3}}^{\dagger}\beta_{k_{4}}^{\dagger}\\
\\
\mathcal{T}_{3} & = & \sum_{k_{1}k_{2}k_{3}k_{4}k_{5}k_{6}}\theta_{k_{1}k_{2}k_{3}k_{4}k_{5}k_{6}}\beta_{k_{1}}^{\dagger}\beta_{k_{2}}^{\dagger}\beta_{k_{3}}^{\dagger}\beta_{k_{4}}^{\dagger}\beta_{k_{5}}^{\dagger}\beta_{k_{6}}^{\dagger}.
\end{array}\label{eq:ClusterOperators}
\end{equation}
The $\theta_{k_{1}k_{2}\ldots}\in\mathbb{C}$ are variational parameters
which are fully antisymmetric such that $\theta_{k_{1}k_{2}\ldots}=\left(-1\right)^{\xi\left(P\right)}\theta_{P\left(k_{1}k_{2}\ldots\right)}$,
where $\xi\left(P\right)$ is the signature of the permutation $P$.
The BUCC ansatz is defined as
\begin{equation}
\left|\Psi\left(\Theta\right)\right\rangle =e^{i\left(\mathcal{T}\left(\Theta\right)+\mathcal{T}^{\dagger}\left(\Theta\right)\right)}\left|\Phi_{0}\right\rangle .\label{eq:UnitaryVariationalAnsatz}
\end{equation}
where $\Theta$ corresponds to the set of variational parameters $\theta_{k_{1}k_{2}\ldots}$
and $\left|\Phi_{0}\right\rangle $ is a reference state. Since the
transformation is unitary $\left|\left\langle \Psi\left(\Theta\right)|\Psi\left(\Theta\right)\right\rangle \right|=1$, $|\Psi\left(\Theta\right)\rangle$ is always normalized. The BUCC ansatz is said to be over single (BUCCS)
or double excitations (BUCCSD) if the cluster operator $\mathcal{T}$
is truncated at the first or second order. 

To variationally optimize the BUCC ansatz, we aim to find the angles $\Theta$ that
minimize the energy
\begin{equation}
\min_{\Theta}E\left(\Theta\right)=\left\langle \Psi\left(\Theta\right)\right|H\left|\Psi\left(\Theta\right)\right\rangle \label{eq:VariationalEnergyProblem}
\end{equation}
subject to the constraint that the number of particles 
\begin{equation}
\begin{array}{rcl}
\left\langle N\left(\varTheta\right)\right\rangle  & = & \left\langle \Psi\left(\Theta\right)\right|N\left|\Psi\left(\Theta\right)\right\rangle \\
\\
 & = & \sum_{p=1}^{M}\left\langle \Psi\left(\Theta\right)\right|a_{p}^{\dagger}a_{p}\left|\Psi\left(\Theta\right)\right\rangle 
\end{array}\label{eq:ParticleNumber}
\end{equation}
should be kept constant, as the quasiparticles operators generally do not preserve
the total particle number. In the next subsection we will
explicitly show how to compute the reference state from the generalized
Hartree-Fock theory before describing the details of the implementation
of the quantum algorithm.

\subsection{Generalized Hartree-Fock theory\label{subsec:GHFT}}

Here we show how to obtain the Bogoliubov matrix \eqref{eq:BogoliubovMatrix}
used to define the reference state \eqref{eq:HFBGroundState}. The
method relies on the theory of fermionic Gaussian states \citep{Kraus09,Kraus10}
for which we review the formalism and a method to obtain the covariance
matrix of the ground state without a self-consistent loop. Fermionic
Gaussian states are a useful starting point for quantum simulations
as they include the family of Slater determinants from Hartree-Fock
theory and Bardeen-Cooper-Schrieffer (BCS) states found in the mean-field theory of superconductivity \cite{Cooper56,Bardeen57}
and can be easily prepared on a quantum computer \citep{Verstraete09}. 

For $M$ fermionic modes, it is convenient to define the $2M$ Majorana
operators 
\begin{equation}
\begin{array}{rcccl}
\gamma_{j} & = & \gamma_{j}^{A} & = & a_{j}^{\dagger}+a_{j}\\
\\
\gamma_{j+M} & = & \gamma_{j}^{B} & = & -i\left(a_{j}^{\dagger}-a_{j}\right)
\end{array}\label{eq:MajoranaOperators}
\end{equation}
as the fermionic analogues of position and momentum operators. Let's
note that we used either the extended index notation (from $1$ to
$2M$) or the $A$,$B$ superscript notation interchangeably throughout
the paper to make the equations clearer. Their commutation relation
satisfies $\left\{ \gamma_{k},\gamma_{l}\right\} =2\delta_{kl}$ such
that $\gamma_{k}^{2}=1$. It is useful to define the vector notation
$\vec{\gamma}^{\top}=\left(\gamma_{1},\ldots,\gamma_{M},\gamma_{M+1},\ldots,\gamma_{2M}\right)$
and write $\vec{\gamma}=\varOmega\vec{a}$ where 
\begin{equation}
\varOmega=\left(\begin{array}{cc}
\mathbf{1} & \mathbf{1}\\
i\mathbf{1} & -i\mathbf{1}
\end{array}\right).\label{eq:OmegaMatrix}
\end{equation}
In this case, $\mathbf{1}$ is the $M\times M$ identity matrix. A
general fermionic Gaussian state \citep{Kraus09} has the form of
the exponential of a quadratic product of fermionic operators
\begin{equation}
\rho=\frac{1}{Z}e^{-\frac{i}{4}\vec{\gamma}^{\top}G\vec{\gamma}},\label{eq:GaussianState}
\end{equation}
where $Z$ is the normalization factor and $G$ is a real and antisymmetric
matrix such that $G^{\top}=-G$. It can be fully characterized by
a real and antisymmetric covariance matrix which is defined by
\begin{equation}
\varGamma_{kl}=\frac{i}{2}\textrm{tr}\left(\rho\left[\gamma_{k},\gamma_{l}\right]\right),\label{eq:CovarianceMatrix}
\end{equation}
where $\left[\cdot,\cdot\right]$ is the commutator.
For a pure Gaussian state, $\varGamma^{2}=-\boldsymbol{1}$, where
$\boldsymbol{1}$ is the $2M\times2M$ identity matrix. In general,
the purity is given by $\chi=-\frac{1}{2M}\textrm{tr}\left(\varGamma^{2}\right)$.
In order to extract $\mathcal{U}$ given a covariance matrix $\varGamma$,
we make use of the complex covariance matrix representation 
\begin{equation}
\varGamma_{c}=\frac{1}{4}\varOmega^{\dagger}\varGamma\varOmega^{\ast}=\left(\begin{array}{cc}
\mathbf{Q} & \mathbf{R}\\
\mathbf{R}^{\ast} & \mathbf{Q}^{\ast}
\end{array}\right),\label{eq:ComplexCovarianceMatrix}
\end{equation}
where $\mathbf{Q}_{kl}=\frac{i}{2}\left\langle \left[a_{k},a_{l}\right]\right\rangle $
and $\mathbf{R}_{kl}=\frac{i}{2}\left\langle \left[a_{k},a_{l}^{\dagger}\right]\right\rangle $ (expectation values are defined as $\left\langle O\right\rangle=\textrm{tr}\left(O\rho\right)$). 
From there, we can define the single-particle density operators $\mathbf{\kappa}\equiv-i\mathbf{Q}$
and $\mathbf{\varrho}\equiv\frac{1}{2}\mathbf{1}-i\mathbf{R}^{\top}$
and recast the Gaussian state in the form of a single-particle density
matrix
\begin{equation}
\mathcal{M}=\left(\begin{array}{cc}
\mathbf{\varrho} & \mathbf{\kappa}^{\dagger}\\
\mathbf{\kappa} & \mathbf{1}-\mathbf{\varrho}^{\top}
\end{array}\right)\label{eq:SingleParticleDensity}
\end{equation}
such that $\mathcal{M}^{2}=\mathcal{M}$ for pure states \citep{Bloch62}.
If we define the matrix $\mathcal{E}=\left(\begin{array}{cc}
\mathbf{0} & \mathbf{0}\\
\mathbf{0} & \mathbf{1}
\end{array}\right)$, then it is possible to find the Bogoliubov transformation \eqref{eq:BogoliubovMatrix}
with the eigenvalue equation 
\begin{equation}
\mathcal{M}\mathcal{U}^{\dagger}=\mathcal{E}\mathcal{U}^{\dagger}.\label{eq:BogoliubovEigenEquation}
\end{equation}
Next, we show how to compute the covariance matrix \eqref{eq:CovarianceMatrix}
approximating the ground state of the Hamiltonian \eqref{eq:Hamiltonian}.

\subsubsection{Finding the ground state\label{subsec:GHFTGroundState}}

These steps are a review of the method found in \citep{Kraus10} aimed
at calculating the covariance matrix approximating the ground state
of an interacting Hamiltonian without a self-consistent loop.

The Hamiltonian \eqref{eq:Hamiltonian} can be rewritten with Majorana
operators in the form 
\begin{equation}
\begin{array}{rcl}
H & = & i\sum_{pq}T_{pq}\gamma_{p}\gamma_{q}\\
\\
 &  & +\sum_{pqrs}V_{pqrs}\gamma_{p}\gamma_{q}\gamma_{s}\gamma_{r}\\
\\
 &  & +i\sum_{pqrstu}W_{pqrstu}\gamma_{p}\gamma_{q}\gamma_{r}\gamma_{u}\gamma_{t}\gamma_{s},
\end{array}\label{eq:HamiltonianMajorana}
\end{equation}
where $T^{\top}=-T$ and $V$ and $W$ are antisymmetric under the
exchange of any two adjacent indices. Expectation values over gaussian
states can be efficiently calculated using Wick's theorem which has
the form 
\begin{equation}
i^{p}\textrm{tr}\left(\rho\gamma_{j_{1}}\ldots\gamma_{j_{2p}}\right)=\textrm{Pf}\left(\varGamma|_{j_{1}\ldots j_{2p}}\right),\label{eq:WicksTheorem}
\end{equation}
where $1\leq j_{1}<\ldots<j_{2p}\leq2M$, $\varGamma|_{j_{1}\ldots j_{2p}}$
is the corresponding submatrix of $\varGamma$ and 
\begin{equation}
\begin{array}{rcl}
\textrm{Pf}\left(\varGamma\right) &=& \frac{1}{2^M M!}\sum_{s\in S_{2M}}{\textrm{sgn}\left(s\right)\prod_{j=1}^{M}\varGamma_{s\left(2j-1\right),s\left(2j\right)}}\\
\\
&=& \sqrt{\textrm{det}\left(\varGamma\right)}
\end{array}
\label{eq:Pfaffian}
\end{equation}
is the Pfaffian of a $2M\times 2M$ matrix defined from the symmetric group $S_{2M}$ where $\textrm{sgn}\left(s\right)$ is the signature of the permutation $s$. Assuming that Wick's theorem holds, we can write an
effective but state dependent quadratic Hamiltonian 
\begin{equation}
h\left(\Gamma\right)=T+6\textrm{tr}_{B}\left(V\Gamma\right)+45\textrm{tr}_{C}\left(W\Gamma\Gamma\right),\label{eq:QuadraticHamiltonian}
\end{equation}
where $\textrm{tr}_{B}\left(V\Gamma\right)_{ij}=\sum_{kl}V_{ijkl}\Gamma_{lk}$
and $\textrm{tr}_{C}\left(W\Gamma\Gamma\right)_{ij}=\sum_{klmn}W_{ijklmn}\Gamma_{kn}\Gamma_{ml}$.
To get the covariance matrix of the reference state, we use the imaginary
time evolution starting from a pure state $\Gamma\left(0\right)^{2}=-\boldsymbol{1}$:
\begin{equation}
\Gamma\left(\tau\right)=O\left(\tau\right)\Gamma\left(0\right)O\left(\tau\right)^{\top},\label{eq:ImaginaryTimeEvolution}
\end{equation}
where the orthogonal time evolution operator is given by
\begin{equation}
O\left(\tau\right)=\mathbb{T}e^{2\int_{0}^{\tau}d\tau'\left[h\left(\Gamma\left(\tau'\right)\right),\Gamma\left(\tau'\right)\right]},\label{eq:TimeEvolutionOperator}
\end{equation}
with $\mathbb{T}$ being the time ordering. The steady state is reached
when $\left[h\left(\Gamma\right),\Gamma\right]=0$. This is guaranteed
to lower the energy of an initial state and keep the purity of the
initial $\Gamma\left(0\right)$ but the imaginary time evolution may
get stuck in a local minimum. A second complementary approach consists
in minimizing the free energy of \eqref{eq:HamiltonianMajorana}.
The procedure simply involves fixed point iterations on the transcendental
equation 
\begin{equation}
\varGamma=\lim_{\beta\rightarrow\infty}\tanh\left[2i\beta h\left(\Gamma\right)\right].\label{eq:FixedPointEvolution}
\end{equation}
In our numerical experiments, we find that an imaginary time evolution
\eqref{eq:ImaginaryTimeEvolution} followed by a fixed point evolution
\eqref{eq:FixedPointEvolution} is numerically stable and consistently
reaches the desired GHF ground state. In the following subsection,
we will show how the theory of matchgates can be used to prepare a
pure Gaussian state on a quantum computer as a reference state for
a variational procedure.

\subsection{The quantum subroutine\label{subsec:QuantumAlgorithm}}

It is expected that quantum computer will enable the simulation of
quantum systems beyond the reach of classical computers. An important
challenge for practical simulations is to prepare the ground state
of interesting Hamiltonians with high accuracy. The VQE protocol \citep{Peruzzo14,McClean16,Wecker15b,OMalley15,Kandala17}
suggests a general procedure to reach this ground state. However,
current implementations of the protocol have to trade long circuit
depth for accuracy in a non-controllable manner. In this subsection,
we introduce a composable VQE ansatz which is both accurate and hardware
efficient with the added advantage of being able to represent states
with BCS-like pairing correlations. Our method relies on the theory
of matchgates and its relation to fermionic linear optics \citep{Valiant02,Terhal02,Bravyi02,Jozsa08,Verstraete09,Brod16}
to both prepare a reference Gaussian state and parametrize an ansatz
with a transformation analogous to fermionic non-linear optics. After
a brief review of the theory of matchgates, we show how a given pure
Gaussian state can be prepared on a quantum register with a linear-depth
algorithm. A different algorithm with the same scaling was recently proposed in \cite{Jiang17}. Unlike the procedure in \cite{Jiang17}, that relies on a gate decomposition strategy, our method has a fixed circuit structure with variable parameters. We then proceed to introduce a low-depth circuit ansatz
with inherited properties of the BUCC ansatz and the apparent accuracy
of the full configuration interaction method.

\subsubsection{Matchgate decomposition of a Bogoliubov transformation\label{subsec:MatchgateDecomposition}}

\begin{figure*}
\begin{centering}
\includegraphics[height=0.8\textheight]{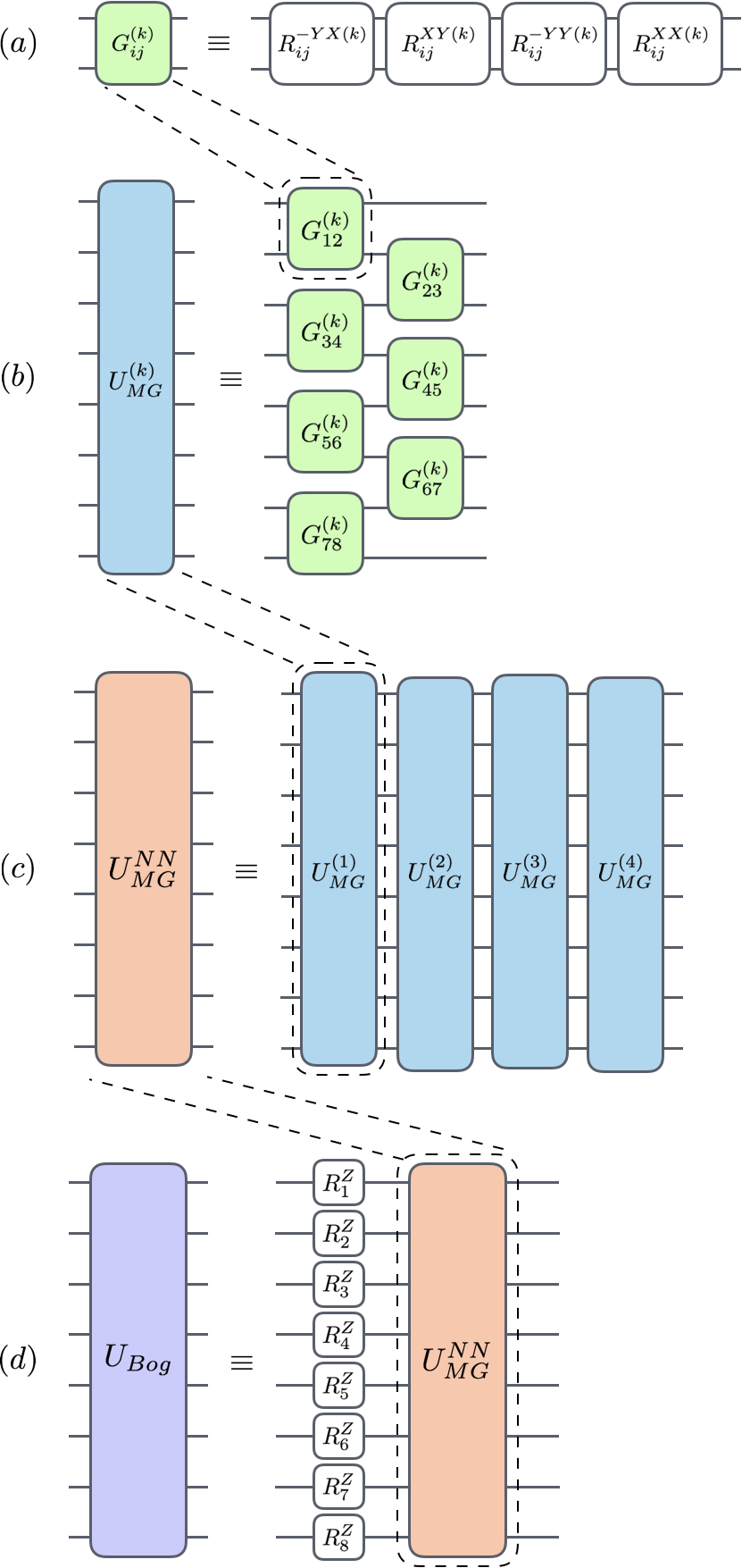}\caption{Example on 8 qubits of the decomposition of $U_{\mathrm{Bog}}$ in
a circuit of local phase rotations and nearest-neighbor matchgates.
In (a), each $G_{ij}^{\left(k\right)}$ is a 2-local operation between
qubits $i$ and $j$ composed of 4 rotations for a layer $k$. As
shown in (b), the unitary $U_{\mathrm{MG}}^{\left(k\right)}$ for
each layer $k$ is built by operating $G_{ij}^{\left(k\right)}$'s
in parallel first on the even pairs of qubits and then on the odd
pairs. Then in (c), the complete sequence of nearest-neighbor matchgates
$U_{\mathrm{MG}}^{\mathrm{NN}}$ is composed by a sequence of $\left\lceil \frac{M}{2}\right\rceil $
layers. In (d), single qubit phase rotations $R_{j}^{Z}$ are used
to complete the $U_{\mathrm{Bog}}$ circuit.\label{fig:UBogDecomposition}}
\par\end{centering}
\end{figure*}

In the computational basis of a 2-qubit Hilbert space, matchgates
\citep{Valiant02} have the general form
\begin{equation}
\mathcal{G}\left(A,B\right)=\left(\begin{array}{cccc}
p & 0 & 0 & q\\
0 & w & x & 0\\
0 & y & z & 0\\
r & 0 & 0 & s
\end{array}\right),\label{eq:Matchgate}
\end{equation}
where $A=\left(\begin{array}{cc}
p & q\\
r & s
\end{array}\right)$ and $B=\left(\begin{array}{cc}
w & x\\
y & z
\end{array}\right)$ are $SU\left(2\right)$ matrices with the same determinant $\det A=\det B$.
They form a group which is generated by the tensor product of nearest-neighbor
Pauli operators 
\begin{equation}
\begin{array}{rcl}
\sigma_{x}^{j}\otimes\sigma_{x}^{j+1} & = & -i\gamma_{j}^{B}\gamma_{j+1}^{A}\\
\\
\sigma_{x}^{j}\otimes\sigma_{y}^{j+1} & = & -i\gamma_{j}^{B}\gamma_{j+1}^{B}\\
\\
\sigma_{y}^{j}\otimes\sigma_{x}^{j+1} & = & i\gamma_{j}^{A}\gamma_{j+1}^{A}\\
\\
\sigma_{y}^{j}\otimes\sigma_{y}^{j+1} & = & i\gamma_{j}^{A}\gamma_{j+1}^{B}\\
\\
\sigma_{z}^{j}\otimes\mathbb{I}^{j+1} & = & -i\gamma_{j}^{A}\gamma_{j}^{B}\\
\\
\mathbb{I}^{j}\otimes\sigma_{z}^{j+1} & = & -i\gamma_{j+1}^{A}\gamma_{j+1}^{B},
\end{array}\label{eq:MatchgateGenerators}
\end{equation}
which also correspond to the Jordan-Wigner transformed product of
all products of nearest-neighbor Majorana operators, therefore establishing
the connection with fermionic gaussian operations. The Bogoliubov transformation
\eqref{eq:BogoliubovTransformation} can be written as an $SO\left(2M\right)$
transformation of the Majorana operators \eqref{eq:MajoranaOperators}
as $\vec{\gamma}'=\mathcal{R}\vec{\gamma}$, where
\begin{equation}
\mathcal{R}=\left(\begin{array}{cc}
\textrm{Re}\left(\mathbf{U}+\mathbf{V}\right) & -\textrm{Im}\left(\mathbf{U}-\mathbf{V}\right)\\
\textrm{Im}\left(\mathbf{U}+\mathbf{V}\right) & \textrm{Re}\left(\mathbf{U}-\mathbf{V}\right)
\end{array}\right).\label{eq:RealBogoliubovTransformation}
\end{equation}
To implement this transformation on a quantum processor, there exists
a quantum circuit of nearest-neighbor matchgates $U_{\mathrm{Bog}}$
acting on $M$ qubits \citep{Jozsa08} such that
\begin{equation}
U_{\mathrm{Bog}}\gamma_{j}U_{\mathrm{Bog}}^{\dagger}=\sum_{k=1}^{2M}\mathcal{R}_{kj}\gamma_{k}.\label{eq:BogoliubovCircuit}
\end{equation}
An example of such a circuit known as the fermionic fast Fourier transform
is described in \citep{Verstraete09}. In general, the Hoffman algorithm
\citep{Hoffman72} can be used to decompose $U_{\mathrm{Bog}}$ in
$2M\left(M-1\right)$ $SO\left(4\right)$ rotations between pairs
of modes and $M$ $SO\left(2\right)$ local phases. In total, these
$2M^{2}-M$ angles correspond to the same number of quantum gates.
Using the fact that quantum gates can be operated in parallel in a
linear chain of qubits, any transformation $\mathcal{R}$ can be implemented
in circuit depth $8\left\lceil \frac{M}{2}\right\rceil +1$, as detailed
in Figure \ref{fig:UBogDecomposition}.

Since the Hoffman method assumes sequential operations on each pair
of modes, we used an optimal control scheme \citep{Khaneja05,Machnes11}
in $SO\left(2M\right)$ to allow an easy parametrization
of gates acting in parallel. This is generally efficient on a classical
computer since the matchgates only operate on a much smaller subspace
of the full $SU\left(2^{M}\right)$ transformation allowed
on $M$ qubits. The transformation $\mathcal{R}$ can be decomposed
in local and nearest-neighbor mode rotations such that
\begin{equation}
\begin{array}{rcl}
\mathcal{R} & = & \prod_{k=1}^{\left\lceil \frac{M}{2}\right\rceil }\left\{ \prod_{\mu,\nu}\prod_{j\in\mathrm{odd}}r_{j,j+1}^{\mu\nu}\left(\theta_{j,j+1}^{\mu\nu\left(k\right)}\right)\right.\\
\\
 &  & \left.\times\prod_{\mu,\nu}\prod_{j\in even}r_{j,j+1}^{\mu\nu}\left(\theta_{j,j+1}^{\mu\nu\left(k\right)}\right)\right\} \\
\\
 &  & \times\prod_{j=1}^{M}r_{jj}^{AB}\left(\theta_{jj}^{AB}\right),
\end{array}\label{eq:SO2MDecomposition}
\end{equation}
where $\mu,\nu\in\left\{ A,B\right\} $ and $j\in\left\{ 1,\ldots,M\right\} $.
The mode rotations are parametrized by the $2M^{2}-M$ angles $\theta_{ij}^{\mu\nu\left(k\right)}$
\begin{equation}
r_{ij}^{\mu\nu}=e^{2\theta_{ij}^{\mu\nu}h_{ij}^{\mu\nu}}\label{eq:TwoModeRotation}
\end{equation}
with $SO\left(2M\right)$ Hamiltonians
\begin{equation}
h_{ij}^{\mu\nu}=\delta_{i\mu,j\nu}-\delta_{j\nu,i\mu}.\label{eq:SO2MHamiltonian}
\end{equation}
The optimal control method maximizes the fidelity function

\begin{equation}
\Phi=\frac{1}{2M}\mathrm{tr}\left\{ \mathcal{R}_{\mathrm{target}}^{\top}\mathcal{R}\left(\Theta\right)\right\} \label{eq:GRAPECostFunction}
\end{equation}
using the gradient

\begin{equation}
\frac{\partial r_{ij}^{\mu\nu}}{\partial\theta_{kl}^{\alpha\beta}}=2h_{ij}^{\mu\nu}r_{ij}^{\mu\nu}\delta_{\alpha\mu}\delta_{\beta\nu}\delta_{ki}\delta_{lj}.\label{eq:RotationDerivative}
\end{equation}
As shown in Figure \ref{fig:UBogDecomposition} on a 8-qubit example,
this decomposition explicitly translates into a quantum circuit of
single qubit phase-rotations
\begin{equation}
R_{j}^{Z}=e^{i\theta_{ii}^{AB}\sigma_{z}^{i}}\label{eq:ZRotation}
\end{equation}
 and nearest-neighbor matchgates

\begin{equation}
G_{ij}^{\left(k\right)}=R_{ij}^{XX\left(k\right)}R_{ij}^{-YY\left(k\right)}R_{ij}^{XY\left(k\right)}R_{ij}^{-YX\left(k\right)},\label{eq:NearestNeighborMatchgate}
\end{equation}
where each rotation corresponds to
\begin{equation}
\begin{array}{rcl}
R_{ij}^{-YX\left(k\right)} & = & e^{-i\theta_{ij}^{AA\left(k\right)}\sigma_{y}^{i}\otimes\sigma_{x}^{j}}\\
\\
R_{ij}^{XY\left(k\right)} & = & e^{i\theta_{ij}^{BB\left(k\right)}\sigma_{x}^{i}\otimes\sigma_{y}^{j}}\\
\\
R_{ij}^{-YY\left(k\right)} & = & e^{-i\theta_{ij}^{AB\left(k\right)}\sigma_{y}^{i}\otimes\sigma_{y}^{j}}\\
\\
R_{ij}^{XX\left(k\right)} & = & e^{i\theta_{ij}^{BA\left(k\right)}\sigma_{x}^{i}\otimes\sigma_{x}^{j}}.
\end{array}\label{eq:NNRotations}
\end{equation}
Each parallel cycle interleaves gates between even and odd nearest
neighbors

\begin{equation}
U_{\mathrm{MG}}^{\left(k\right)}=\prod_{i\in\mathrm{odd}}G_{i,i+1}^{\left(k\right)}\prod_{i\in\mathrm{even}}G_{i,i+1}^{\left(k\right)}\label{eq:UBogOneLayer}
\end{equation}
and there are $\left\lceil \frac{M}{2}\right\rceil $ cycles in total:
\begin{equation}
U_{\mathrm{MG}}^{\mathrm{NN}}=\prod_{k=1}^{\left\lceil \frac{M}{2}\right\rceil }U_{\mathrm{MG}}^{\left(k\right)}.\label{eq:UBogManyLayers}
\end{equation}
Finally, the unitary Bogoliubov transformation can be composed as

\begin{equation}
U_{\mathrm{Bog}}=U_{\mathrm{MG}}^{\mathrm{NN}}\prod_{i=1}^{M}R_{i}^{Z}\label{eq:UBogAlgo}
\end{equation}
and is also a gaussian operation of the form $U_{\mathrm{Bog}}=e^{i\sum_{pq}\tau_{pq}\gamma_{p}\gamma_{q}}$,
where $\tau^{\top}=-\tau$. In the case where the reference state is a Slater determinant, only number-conserving matchgates are required to prepare the state and the depth of the circuit would scale as $4\left\lceil \frac{M}{2}\right\rceil +1$ (since all $\theta_{ij}^{AA\left(k\right)}$ and $\theta_{ij}^{BB\left(k\right)}$ are set to zero). It should be noticed that a unitary coupled
cluster ansatz truncated at first order $e^{i\left(\mathcal{T}_{1}\left(\Theta\right)+\mathcal{T}_{1}^{\dagger}\left(\Theta\right)\right)}$
is also a gaussian transformation and can be implemented in the same
way as $U_{\mathrm{Bog}}$ with no trotterization. In what follows,
we introduce a VQE scheme that builds on this observation by introducing
non-matchgate variational terms into a gate sequence similar to the
$U_{\mathrm{Bog}}$ decomposition.

\subsubsection{A low-depth circuit ansatz\label{subsec:LinearDepthCircuits}}

The Bogoliubov transformation \eqref{eq:UBogAlgo} acts as a change
of basis of the fermionic modes. Therefore, one can simply follow
the VQE protocol \citep{Peruzzo14} to implement the BUCC ansatz \eqref{eq:UnitaryVariationalAnsatz}
and measure the expectation values $\left\langle \tilde{H}\right\rangle =\left\langle U_{\mathrm{Bog}}HU_{\mathrm{Bog}}^{\dagger}\right\rangle $
and $\left\langle \tilde{N}\right\rangle =\left\langle U_{\mathrm{Bog}}NU_{\mathrm{Bog}}^{\dagger}\right\rangle $
in the modified basis to prepare an approximate ground state of \eqref{eq:Hamiltonian}.
This has the advantage of extending the range of Hamiltonians that
can be processed to those with non-number conserving terms (like pairing
fields) when compared to the traditional unitary coupled cluster ansatz.
However, the change of basis may significantly increase the number
of terms that have to be measured. In order to reduce the number of
measurements in the VQE protocol, one can start in the product
state \eqref{eq:HFBGroundState} and carry out the variational unitary
\eqref{eq:UnitaryVariationalAnsatz} in the quasiparticle basis, followed by an inverse Bogoliubov transformation using matchgates and
measurement of the expectation values of the Hamiltonian \eqref{eq:Hamiltonian}
and the number operator $N$ in the original fermionic orbital basis.
In the quasiparticle basis, we can map the Bogoliubov operators to
qubit operators with the Jordan-Wigner transformation \citep{Jordan28,Ortiz01,Seeley12}
since they follow the canonical anti-commutation relation
\begin{equation}
\begin{array}{rcl}
\beta_{p}^{\dagger} & = & \left(-1\right)^{p-1}\left(\bigotimes_{j=1}^{p-1}\sigma_{z}\right)\otimes\sigma_{+}\\
\\
\beta_{p} & = & \left(-1\right)^{p-1}\left(\bigotimes_{j=1}^{p-1}\sigma_{z}\right)\otimes\sigma_{-}
\end{array}\label{eq:JordanWigner}
\end{equation}
and use the same mapping for Fermionic operators $a_{p}^{\dagger}$
and $a_{p}$ after the Bogoliubov transformation. Still, assuming that
the number of fermionic particles is proportional to the number of
orbitals, a major caveat of BUCCSD-like schemes is that the number
of variational parameters will scale as $O\left(M^{4}\right)$. In
the Jordan-Wigner picture, these terms can be implemented with $O\left(M^{6}\right)$
gates \citep{Hastings15,Romero17}. It is expected that near-term
quantum processor will continue to suffer from error rates that make
this type of scaling impractical, and therefore more hardware-efficient VQE schemes
must be sought \citep{Kandala17}.

\begin{figure*}
\begin{centering}
\includegraphics[height=0.8\textheight]{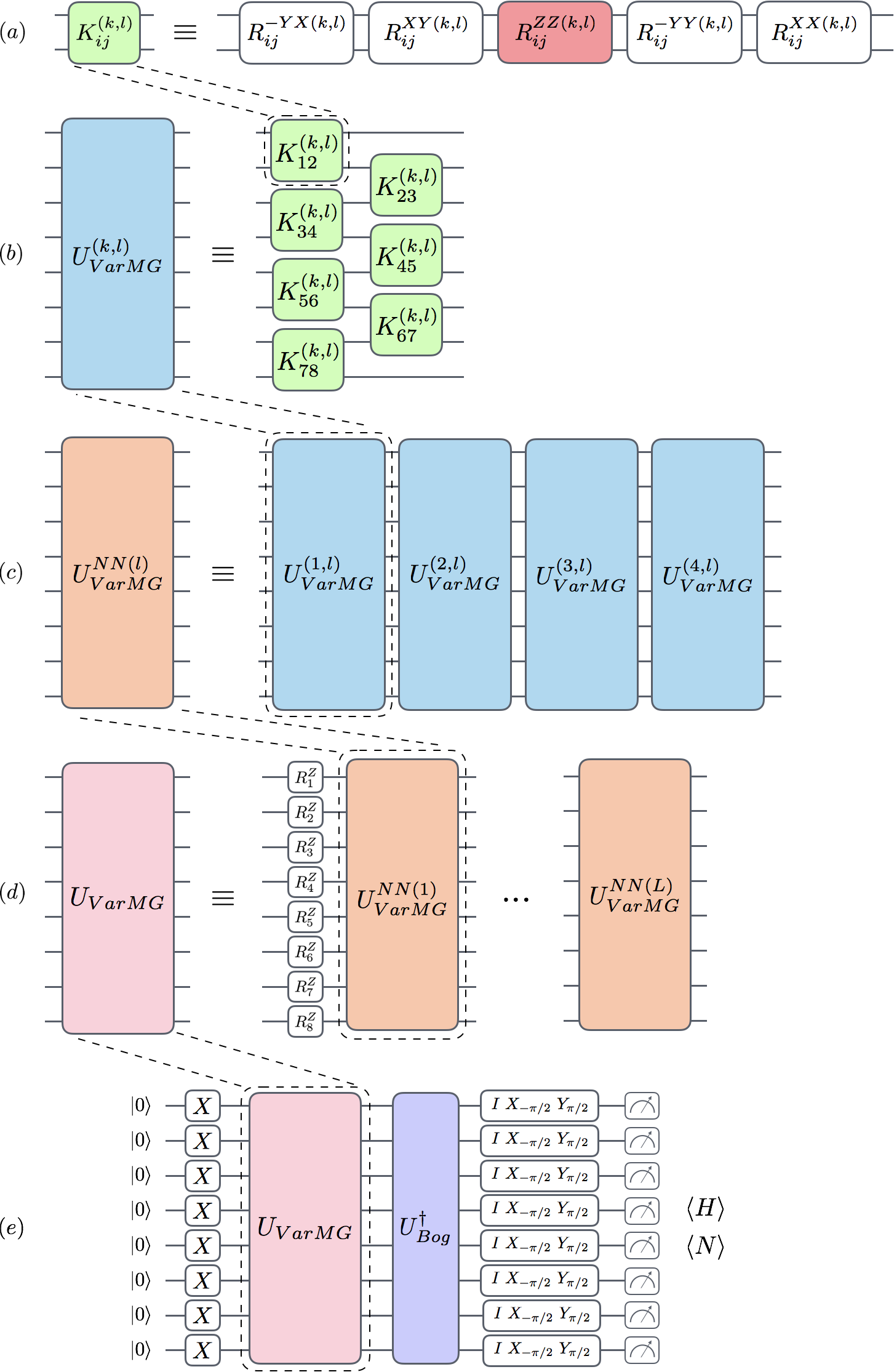}\caption{Gate decomposition of the $L$-cycle LDCA on a linear chain of 8 qubits.
In (a), each $K_{ij}^{\left(k,l\right)}$ is a 2-local operation between
qubits $i$ and $j$ composed of 5 rotations for a layer $k$. In
(b), we build the unitary $U_{\mathrm{VarMG}}^{\left(k,l\right)}$
for each layer $k$ by applying $K_{ij}^{\left(k,l\right)}$'s in
parallel first on the even pairs and then on the odd pairs. In (c),
a cycle $U_{\mathrm{VarMG}}^{\mathrm{NN}\left(l\right)}$ is composed
by a sequence of $\left\lceil \frac{M}{2}\right\rceil $ layers. In
(d), we show the $L$-cycle construction of $U_{\mathrm{VarMG}}$
with one round of variational phase rotations. The full LDCA protocol
is shown in (e) with the initial preparation of the quasiparticle
vacuum and the transformation to the original fermionic basis $U_{\mathrm{Bog}}^{\dagger}$.\label{fig:UVarDecomposition}}
\par\end{centering}
\end{figure*}

Given that the gate decomposition of $U_{\mathrm{Bog}}$ can also
exactly parametrize a BUCCS VQE protocol in linear circuit depth,
we propose using a scheme augmented with nearest-neighbor phase coupling
$\sigma_{z}\otimes\sigma_{z}$ rotations to mimic the effects of the
quartic variational terms of $\mathcal{T}_{2}$. Related ideas have
already been explored in efficient classical non-gaussian variational
methods with great success \citep{Shi17}. In a loose sense, our scheme
is a parametrized fermionic non-linear optics circuit that does not
involve any trotterization of the variational terms. The algorithm
is illustrated in Figure \ref{fig:UVarDecomposition}. As a first
step, the quasiparticle vacuum \eqref{eq:HFBGroundState}
is prepared in the Bogoliubov picture with $X=\left(\begin{array}{cc}
0 & 1\\
1 & 0
\end{array}\right)$ gates acting on each qubits to yield the state $\left|1\right\rangle ^{\otimes M}$
in the computational basis. In what follows, we will define a $L$-cycle
ansatz built from nearest-neighbor variational matchgates augmented
with $\sigma_{z}\otimes\sigma_{z}$ rotations. The measurement of
the expectation values can be done in the original basis by applying
the inverse Bogoliubov transformation $U_{\mathrm{Bog}}^{\dagger}$
defined previously.

In a cycle $l$ of the low-depth circuit ansatz (LDCA), the nearest-neighbor
matchgates \eqref{eq:NearestNeighborMatchgate} are replaced by

\begin{equation}
\begin{array}{rcl}
K_{ij}^{\left(k,l\right)}\left(\Theta_{i,j}^{\left(k,l\right)}\right) & = & R_{ij}^{XX\left(k,l\right)}R_{ij}^{-YY\left(k,l\right)}\\
\\
 &  & \times R_{ij}^{ZZ\left(k,l\right)}R_{ij}^{XY\left(k,l\right)}R_{ij}^{-YX\left(k,l\right)},
\end{array}\label{eq:NearestNeighborVariationalGates}
\end{equation}
where the rotations are defined as
\begin{equation}
\begin{array}{rcl}
R_{ij}^{-YX\left(k,l\right)} & = & e^{-i\theta_{ij}^{-YX\left(k,l\right)}\sigma_{y}^{i}\otimes\sigma_{x}^{j}}\\
\\
R_{ij}^{XY\left(k,l\right)} & = & e^{i\theta_{ij}^{XY\left(k,l\right)}\sigma_{x}^{i}\otimes\sigma_{y}^{j}}\\
\\
R_{ij}^{ZZ\left(k,l\right)} & = & e^{i\theta_{ij}^{ZZ\left(k,l\right)}\sigma_{z}^{i}\otimes\sigma_{z}^{j}}\\
\\
R_{ij}^{-YY\left(k,l\right)} & = & e^{-i\theta_{ij}^{-YY\left(k,l\right)}\sigma_{y}^{i}\otimes\sigma_{y}^{j}}\\
\\
R_{ij}^{XX\left(k,l\right)} & = & e^{i\theta_{ij}^{XX\left(k,l\right)}\sigma_{x}^{i}\otimes\sigma_{x}^{j}}.
\end{array}\label{eq:NNVariationalRotations}
\end{equation}
Each layer $k$ applies those variational rotations in parallel first
on the even pairs and then on the odd pairs such that

\begin{equation}
\begin{array}{rcl}
U_{\mathrm{VarMG}}^{\left(k,l\right)}\left(\Theta^{\left(k,l\right)}\right) & = & \prod_{i\in\mathrm{odd}}K_{i,i+1}^{\left(k,l\right)}\left(\Theta_{i,i+1}^{\left(k,l\right)}\right)\\
\\
 &  & \times\prod_{i\in\mathrm{even}}K_{i,i+1}^{\left(k,l\right)}\left(\Theta_{i,i+1}^{\left(k,l\right)}\right).
\end{array}\label{eq:UVarOneLayer}
\end{equation}
A cycle $l$ is composed of $\left\lceil \frac{M}{2}\right\rceil $
layers such that the variational ansatz is equivalent to a BUCCS transformation
when the $\theta_{ij}^{ZZ\left(k,l\right)}$ are equal to zero:
\begin{equation}
U_{\mathrm{VarMG}}^{\mathrm{NN}\left(l\right)}\left(\Theta^{\left(l\right)}\right)=\prod_{k=1}^{\left\lceil \frac{M}{2}\right\rceil }U_{\mathrm{VarMG}}^{\left(k,l\right)}\left(\Theta^{\left(k,l\right)}\right).\label{eq:UVarManyLayers}
\end{equation}
Finally, the $L$ cycle are assembled sequentially to form the complete
variational ansatz
\begin{equation}
U_{\mathrm{VarMG}}\left(\Theta\right)=\prod_{l=1}^{L}U_{\mathrm{VarMG}}^{\mathrm{NN}\left(l\right)}\left(\Theta^{\left(l\right)}\right)\prod_{i=1}^{M}R_{i}^{Z}\left(\theta_{i}^{Z}\right),\label{eq:UVarCycles}
\end{equation}
with only one round of variational phase rotations
\begin{equation}
R_{i}^{Z}\left(\theta_{i}^{Z}\right)=e^{i\theta_{i}^{Z}\sigma_{z}^{i}}.\label{eq:ZVarRotation}
\end{equation}
 The variational state therefore has the form
\begin{equation}
\left|\Psi\left(\Theta\right)\right\rangle =U_{\mathrm{Bog}}^{\dagger}U_{\mathrm{VarMG}}\left(\Theta\right)\prod_{i=1}^{M}X_{i}\left|0\right\rangle ^{\otimes M},\label{eq:VarAnsatz}
\end{equation}
where it can be noticed that the $L=0$ case is simply equivalent
to producing the GHF state. There are 5 variational angles per $K_{ij}^{\left(k,l\right)}$
and $M-1$ of those terms per layer. Since each cycle has $\left\lceil \frac{M}{2}\right\rceil $
layers, a $L$-cycle circuit has $5L\left(M-1\right)\left\lceil \frac{M}{2}\right\rceil +M$
variational angles, the extra term arising from the round of phase
rotations. Since gates can be operated in parallel in a linear chain
of qubits, the circuit depth is $\left(10L+8\right)\left\lceil \frac{M}{2}\right\rceil +4$ when we account for $U_{\mathrm{Bog}}^{\dagger}$ and the initial round of single-qubit $X$ gates (this includes the final single-qubit rotations, $R_y(\frac{\pi}{2})$ or $R_x(-\frac{\pi}{2})$ gates (or equivalent), to measure the terms of the Hamiltonian in the form of Pauli strings). Therefore, this VQE scheme is hardware
efficient in the sense that the circuit depth is linear in the number
of qubits. The accuracy can also be systematically improved by increasing
the number of cycles until either convergence is reached or errors
dominate the precision of the result.

In the following section, we outline an implementation to compute the analytical gradient of the LDCA using quantum resources, which could be useful during the optimization procedure in VQE by guiding the search for the ground state and its energy.

\subsection{Gradient Evaluation for LDCA\label{subsec:GradientEvaluation}}

% We might need to rethink how to introduce to update this intro
When optimizing the ansatz parameters to minimize the total energy, there may be a need to implement gradients depending on the selected optimization procedure. While direct search algorithms are generally more robust to noise than gradient-based approaches, they may require larger numbers of function evaluations \citep{Kolda06}. On the other hand, numerical implementations of gradients rely heavily on the step size for accuracy. However, step sizes that are too small may lead to numerical instability and higher sampling cost. In addition, implementation of step sizes corresponding to desired accuracy are limited by experimental errors.

An alternative approach that exhibits high accuracy while maintaining reasonable computational cost may be to evaluate the gradient directly on the quantum computer given that the analytical form of the gradient is available. Here we employ a scheme similar to one outlined in \citep{Romero17} tailored to implement the analytical gradient of the LDCA unitary using an extra qubit and controlled two-qubit rotations.
%%%%
Recall the unitary for the complete variational ansatz shown in \eqref{eq:UVarCycles}, which we called $U_{VarMG}(\Theta)$ parametrized by angles $\Theta$. For this derivation, we will ignore the products of Z-rotations in the definition but computing the gradient with respect to these angles should be more straightforward. These initial Z-rotations are not as "nested" within the LDCA framework, so the gradient corresponding to one of such angles, say $\theta_j$, simply involves inserting a controlled-Z gate following the unitary $\text{exp}(-i \theta_j Z)$, to the circuit (where we use an ancilla qubit as the control qubit). Thus, we will instead focus on finding the gradients of the term $\prod_{l=1}^{L} U^{NN}_{VarMG}(\Theta^{(l)})$, which we will call $U_{VarMG}^{'}(\Theta)$.

\begin{figure*}
\begin{centering}
\includegraphics[width=0.6\textheight]{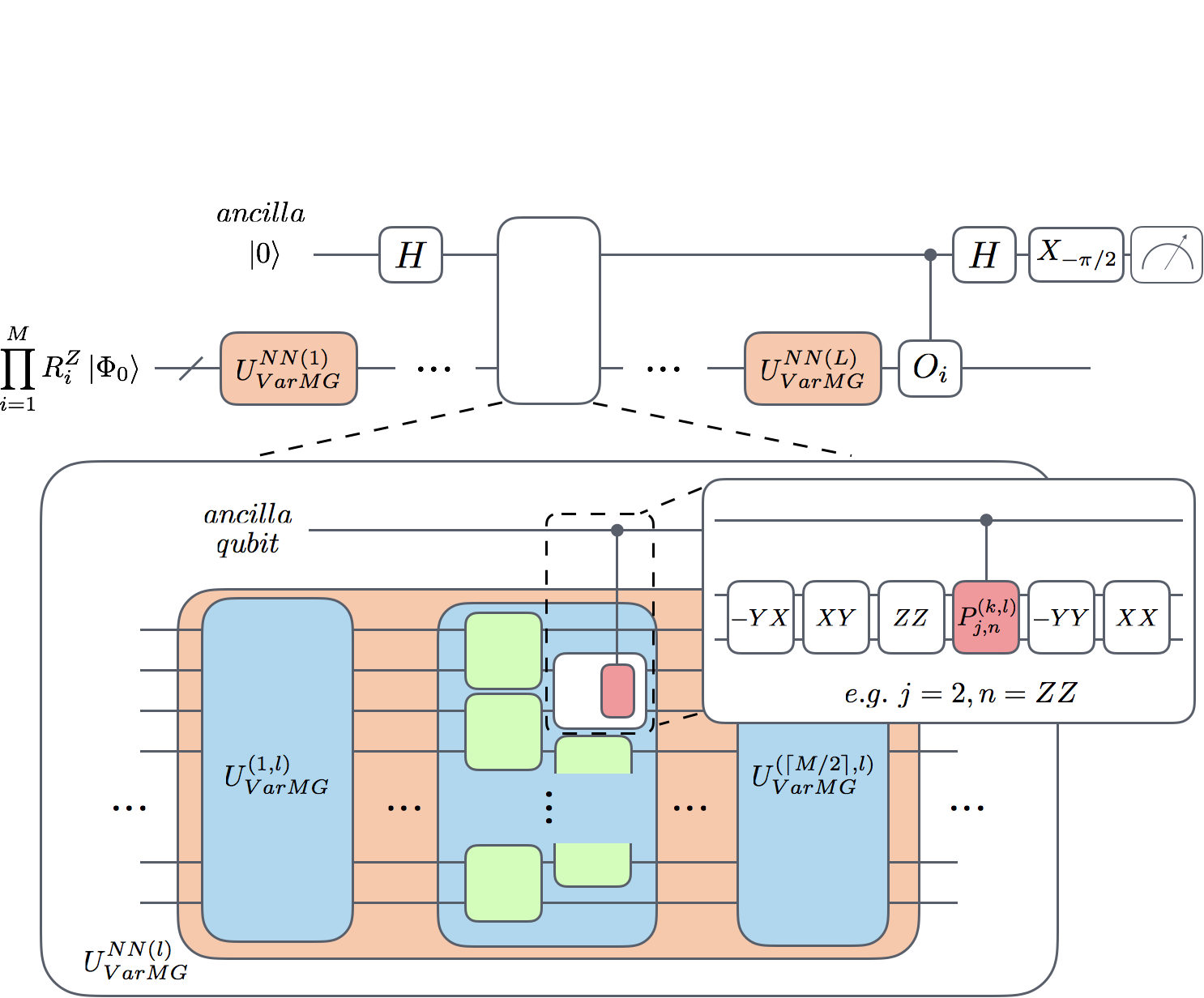}\caption{Circuit using an ancilla qubit to measure the imaginary component of $\langle \Phi_0 | V^{(k, l) \dag}_{j, n} O_i U_{VarMG}^{'} | \Phi_0 \rangle$ required to compute $\frac{\partial E(\Theta)}{\partial \theta_{j, n}^{(k, l)}}$. This figure illustrates an instance of the circuit where $j=2$ and $n=ZZ$.\label{fig:LDCA_gradient_circuit}}
\par\end{centering}
\end{figure*}

Consider the state $\Psi(\Theta)$, prepared by applying $U_{VarMG}(\Theta)$ to $\left|\Phi_0\right\rangle$, where $\left|\Phi_0\right\rangle$ corresponds to a reference state that does not depend on $\Theta$. Here we wish to compute the derivative of the expectation value of the energy $E(\Theta) = \langle \Psi (\Theta)| H | \Psi (\Theta)\rangle$ with respect to each parameter in $\Theta$. We will use the label $\theta_{j, n}^{(k, l)}$ for each parameter where $j$ refers to the index of the qubit in the register, $l$ to the circuit cycle, $k$ to the circuit layer, and $n$ to the appropriate Pauli string (in this case, $n \in \{-YX, XY, ZZ, -YY, XX\}$). Considering a Hamiltonian $H$ that is independent of $\Theta$, the derivative with respect to $\theta_{j, n}^{(k, l)}$ is given by

\begin{subequations}\label{eq:energyGradient}
\begin{align}
\frac{\partial E(\Theta)}{\partial \theta_{j, n}^{(k, l)}} & =  \langle \Phi_0 | U^\dag \  H \ \frac{\partial U}{\partial \theta_{j, n}^{(k, l)}} | \Phi_0 \rangle
+ \langle \Phi_0 | \frac{\partial U^\dag}{\partial \theta_{j, n}^{(k)}} \  H \ U | \Phi_0 \rangle \label{eq:energyGradientA}\\
& =  i \Big( \langle \Phi_0 | U^\dag \  H \ V^{(k, l)}_{j, n} | \Phi_0 \rangle - \langle \Phi_0 | V^{(k, l) \dag}_{j, n} \  H \ U | \Phi_0 \rangle \Big) \label{eq:energyGradientB}\\
& =  2 \ \text{Im} \Big(\langle \Phi_0 | V^{(k, l) \dag}_{j, n} \  H \ U | \Phi_0 \rangle \Big) \label{eq:energyGradientC}
\end{align}
\end{subequations}

\noindent where the operator $V^{(k, l)}_{j, n}(\Theta)$ is nearly identical to the unitary $U^{'}_{VarMG}$ except with a string of Pauli matrices $P^{k,l}_{j,n}$ inserted after the rotation term $R^{n(k, l)}_{j, j+1} = \text{exp}(i \theta^{k,l}_{j,n} P^{k,l}_{j,n})$ included in the nearest-neighbor matchgate term $K^{(k,l)}_{j, j+1}$ and so on. 

% % Clean up this equation!
% \begin{equation}
% \begin{array}{rcl}

% V^{(k, l)}_{j, n} = U^{NN(1)}_{VarMG} \cdot \cdot \cdot \\
% (U^{(1, l)}_{VarMG} \cdot \cdot \cdot \\
% (K^{k, l}_{1,2} \cdot \cdot \cdot K^{k, l}_{j, j+1} P^{k, l}_{j, n} \cdot \cdot \cdot K^{k,l}_{M-1, M}) \\
% \cdot \cdot \cdot U^{(\left\lceil \frac{M}{2}\right\rceil, l)}_{VarMG}) \\
% \cdot \cdot \cdot U^{NN(L)}_{VarMG}

% \end{array}\label{eq:GradientInsertedTerm}
% \end{equation}

To compute the expectation value of the energy, we can employ the Hamiltonian averaging procedure \citep{McClean14, McClean16}. This involves measuring the expectation value of every term in the Hamiltonian and summing over them as shown in \eqref{eq:HamAveraging}. Note that each term, which we call $O_i$, is a product of Pauli matrices obtained by performing the Jordan-Wigner or Bravyi-Kitaev transformation on the corresponding term in the second quantized Hamiltonian from \eqref{eq:Hamiltonian}.

\begin{equation}
\begin{array}{rcl}

E = \sum_i h_i \langle O_i \rangle.

\end{array}\label{eq:HamAveraging}
\end{equation}

% label each equation in eq:energyGradient
\noindent Substituting \eqref{eq:HamAveraging} into \eqref{eq:energyGradientC}, the gradient can be expressed as:

\begin{equation}
\begin{array}{rcl}

\frac{\partial E(\Theta)}{\partial \theta_{j, n}^{(k, l)}} & = & 2 \sum_i \ h_i \ \text{Im} \Big(\langle \Phi_0 | V^{(k, l) \dag}_{j, n}(\Theta) O_i U(\Theta) | \Phi_0 \rangle \Big)

\end{array}\label{eq:GradientFinalEqn}
\end{equation}

%\begin{equation}
%\begin{array}{rcl}

%\frac{|0\rangle \otimes (U|\Phi_0 \rangle + O_i V^{(k, l) \dag}_{j, n} |\Phi_0 \rangle) + |1\rangle \otimes (U|\Phi_0 \rangle - O_i V^{(k, l) \dag}_{j, n} |\Phi_0 \rangle)}{2}.

%\end{array}\label{eq:GradientCircuitFinalState}
%\end{equation}

\noindent Each of the terms in the sum above can be computed using the circuit shown in Figure \ref{fig:LDCA_gradient_circuit}. For a practical physical implementation of the analytical gradient, a circuit layout similar to one highlighted in \citep{DDW16b} could be used, in which the control qubit of the gradient circuit is connected to all qubits in the register.

In the following section, we numerically benchmark the BUCC ansatz and
LDCA on small instances of the Fermi-Hubbard model and the
automerization reaction of cyclobutadiene, where we find that LDCA
is able to prepare the exact ground state of those systems.

\section{Numerical examples\label{sec:Numerics}}

In this section, we numerically test the performance of the previously
described algorithms on instances of strongly correlated systems in
condensed matter and quantum chemistry. Specifically, in subsection \ref{subsec:FermiHubbard}, we analyze the
behavior of the ansatz on the Fermi-Hubbard model at half-filling
at different interaction strengths.
In subsection \ref{subsec:Molecule}, we study the automerization reaction of cyclobutadiene modeled using the Pariser-Parr-Pople (PPP) Hamiltonian \citep{Pariser53a,Pariser53b,Pople53}.
In both cases, the Hamiltonians are mapped to 8-qubit registers and
we compare the energy and wavefunction accuracies for approximating the exact ground state for the following methods ansatzes: GHF, BUCCSD, and LDCA with 1 and 2 cycles. 

In these cases, the state initialization
has 8 single qubit $X$ gates operated in parallel and the inverse
Bogoliubov transformation has one layer of single qubit phase rotations
and 112 nearest-neighbor matchgates. The state initialization and
$U_{\mathrm{Bog}}^{\dagger}$ circuit add up to a circuit depth of
34. The LDCA method adds a layer of variational phase rotations and
140 nearest-neighbor gates per cycle. Therefore 1-cycle LDCA adds
41 to the circuit depth (for a total of 75 with 148 variational parameters)
and 2-cycle LDCA adds 81 to the circuit depth (for a total of 115
with 288 variational parameters). 

For the numerical examples presented here,
we find that 2-cycle LDCA is able to exactly recover the ground state
of the simulated systems while 1-cycle LDCA performs better than the
GHF solution but is not as accurate as BUCCSD. An important caveat
is that the 2-cycle LDCA has more variational parameters (288) than the
dimensions of the Hilbert space ($2^{8}=256$) but we still consider
the result relevant for experimental implementations since the depth
of the circuit is much shorter than what could be achieved with BUCC
up to fourth order, which is required to recover the exact ground state
of systems studied.

\subsection{Fermi-Hubbard model\label{subsec:FermiHubbard}}

The Fermi-Hubbard model \citep{Hubbard63} is a prototypical example
of correlated electrons. It is described by a tight-binding lattice
of electrons interacting through a local Coulomb force. The Hamiltonian
is given by
\begin{equation}
\begin{array}{rcl}
H^{\textrm{FH}} & = & -t\sum_{\left\langle p,q\right\rangle }\sum_{\sigma=\uparrow,\downarrow}\left(a_{p\sigma}^{\dagger}a_{q\sigma}+a_{q\sigma}^{\dagger}a_{p\sigma}\right)\\
\\
 &  & -\mu\sum_{p}\sum_{\sigma=\uparrow,\downarrow}\left(n_{p\sigma}-\frac{1}{2}\right)\\
\\
 &  & +U\sum_{p}\left(n_{p\uparrow}-\frac{1}{2}\right)\left(n_{p\downarrow}-\frac{1}{2}\right),
\end{array}\label{eq:FermiHubbardHamiltonian}
\end{equation}
where $t$ is the kinetic energy between nearest-neighbor sites $\left\langle p,q\right\rangle $,
$U$ is the static Coulomb interaction and $\mu$ is the chemical
potential. The number operator is $n_{p\sigma}=a_{p\sigma}^{\dagger}a_{q\sigma}$.
While the one-dimensional Fermi-Hubbard model can be solved exactly
with the Bethe ansatz \citep{Lieb03,Essler05}, the two-dimensional
version can only be solved exactly for very specific values of the
parameters and a general solution remains elusive. The phase diagram
of the 2D model is known to be very rich and there are strong arguments
that a better understanding of the model could yield the key
to explain the physics of high-temperature cuprate superconductors
\citep{Anderson87,Leggett99,Anderson04}.

\begin{figure*}
\begin{centering}
\includegraphics[width=0.8\textwidth]{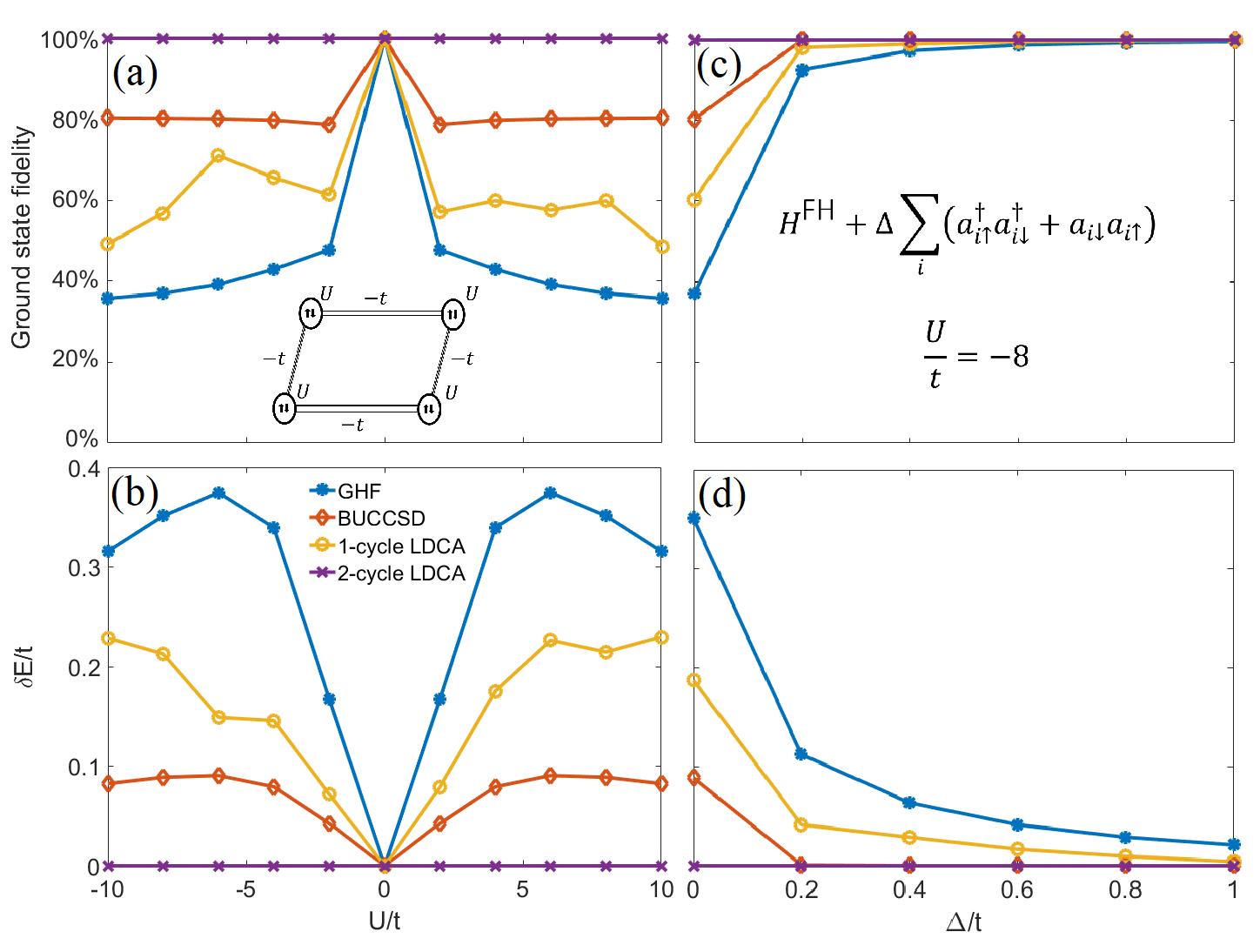}\caption{In (a), we show the fidelity of the ground state preparation of a
$2\times2$ cluster of the Fermi-Hubbard model as a function of the
interaction parameter $U$. The energy difference with the exact ground
state with respect to the various methods is shown in (b). The energies
are normalized by the hopping term $t$. In (c) and (d), we show respectively the fidelity and the energy difference in the case of an attractive cluster $\frac{U}{t}=-8$ with an additional s-wave pairing term $\Delta$.\label{fig:Hubbard}}
\par\end{centering}
\end{figure*}

Hybrid quantum-classical methods to systematically approximate the
phase diagram of the Fermi-Hubbard model in the thermodynamical limit
are known \citep{DDW16,DDW16b} but they require preparing the ground
state of a large cluster of the model with an accuracy that cannot
be reached by previously proposed methods \citep{Wecker15}. Here, we investigate the performance of the ansatz detailed in section
\ref{sec:GVQE} on an example of a $2\times2$ cluster of the Fermi-Hubbard
model at half-filling ($\mu=0$) that can be implemented on a 8-qubit
quantum processor. As shown in figure \ref{fig:Hubbard}, the GHF
method performs well for small values of the interaction strength
$\frac{U}{t}$ and exactly describes the tight-binding case where
the Hamiltonian is quadratic. The BUCCSD ansatz offers a significant
improvement over the GHF solution but fails to reach the exact ground
state at strong interaction strengths. While 1-cycle LDCA offers
an intermediate solution between GHF and BUCCSD, the 2-cycle LDCA
solution performs surprisingly well as it is able to reach the exact
ground state up to numerical accuracy for all values of the interaction strength. In all cases
the preparation fidelity $\left|\left\langle \Psi\left(\Theta\right)|\Psi_{0}\right\rangle \right|^{2}$
is directly correlated with the energy difference $\delta E$ between
the prepared state and the exact ground state $\left|\Psi_{0}\right\rangle $. We also show that all methods are able to handle Hamiltonians with pairing terms by introducing an artificial $\Delta\sum_{i}\left(a_{i\uparrow}^{\dagger}a_{i\downarrow}^{\dagger}+a_{i\downarrow}a_{i\uparrow}\right)$. The accuracy of all methods improves with increasing $\frac{\Delta}{t}$ as the ground state gets closer to a fermionic Gaussian state.

We also tested a simpler one dimensional cluster of the Fermi-Hubbard
model with 2 sites and found that it was possible to reach the exact
ground state with both BUCCSD and the 1-cycle LDCA method for all
values of the parameter $U$. This is expected for the BUCCSD method
as this is equivalent to a full configuration interaction parametrization
in this specific case. We do not have sufficient information to determine
the number of cycles $L$ required by LDCA to reach the ground state
as a function of the cluster size since it would require much more
intense numerics. However, the fact that a $2\times1$ cluster requires
only 1 cycle and that the $2\times2$ case reaches the ground state
in 2 cycles leave open the possibility that the scaling is not an
exponential function of the cluster size.

\subsection{Cyclobutadiene\label{subsec:Molecule}}

As an example of a quantum chemistry application, we studied the accuracy of the proposed methods in the description of cyclobutadiene automerization. The study of this reaction has been particularly challenging for theoretical chemists due to the strongly correlated character of the open-shell $D_{4h}$ transition state in contrast with the weakly correlated character of the closed-shell $D_{2h}$ ground state ($^{1}A_{1g}$) \citep{Szalay12}. An accurate theoretical treatment of the transition state would allow to confirm several observations about the mechanism, such as the alleged change in the aromatic character of the molecule between its ground and transition states as well as the involvement of a tunneling carbon atom in the reaction \citep{Szalay12,Arnold13,Arnold91,Arnold93}. In addition, it would serve as a confirmation of the energy barrier for the automerization, for which experimental reports vary between $1.6$ and $12.0$ kcal/mol \citep{Whitman82}.

Although the Hamiltonian for cyclobutadiene can be obtained from a Hartree-Fock or a Complete Active Space (CAS) standard quantum chemistry calculation, we opted to describe the reaction using a Pariser-Parr-Pople (PPP) model Hamiltonian \citep{Pariser53a,Pariser53b,Pople53}. The PPP model captures the main physics of $\pi$-electron
systems such as cyclobutadiene and also establishes a direct connection to the Fermi-Hubbard Hamiltonian studied in the previous section. Using this model, the Hamiltonian of cyclobutadiene can be written as
\begin{equation}
\begin{array}{rcl}
H^{\text{PPP}} & = & \sum_{i<j}t_{ij}E_{ij}\\
\\
 &  & +\sum_{i}U_{i}n_{i\alpha}n_{i\beta}+V_{c}\\
\\
 &  & +\frac{1}{2}\sum_{ij}\gamma_{ij}(n_{i\alpha}+n_{i\beta}-1)(n_{j\alpha}+n_{j\beta}-1),
\end{array}\label{eq:PPPHamiltonian}
\end{equation}
where $E_{ij}=\sum_{\sigma=\alpha,\beta}a_{i\sigma}^{\dagger}a_{j\sigma}+a_{j\sigma}^{\dagger}a_{i\sigma}$,
$n_{i\sigma}=a_{i\sigma}^{\dagger}a_{i\sigma}$, and the variables $\gamma_{ij}$
are parameterized by the Mataga-Nishimoto formula \citep{mataga1957}
\begin{equation}
\gamma_{ij}(r_{ij})=\frac{1}{1/U+r_{ij}}.\label{eq:MatagaNishimotoFormula}
\end{equation}
The $t_{ij}$, $U$, and $V_{c}$ parameters were obtained from \citep{Schmalz11,Schmalz13}
as a function of the dimensionless reaction coordinate, $\lambda$, and the geometries of
the ground as well as transition states were optimized at this level
of theory.

\begin{figure}
\begin{centering}
\includegraphics[width=0.9\columnwidth]{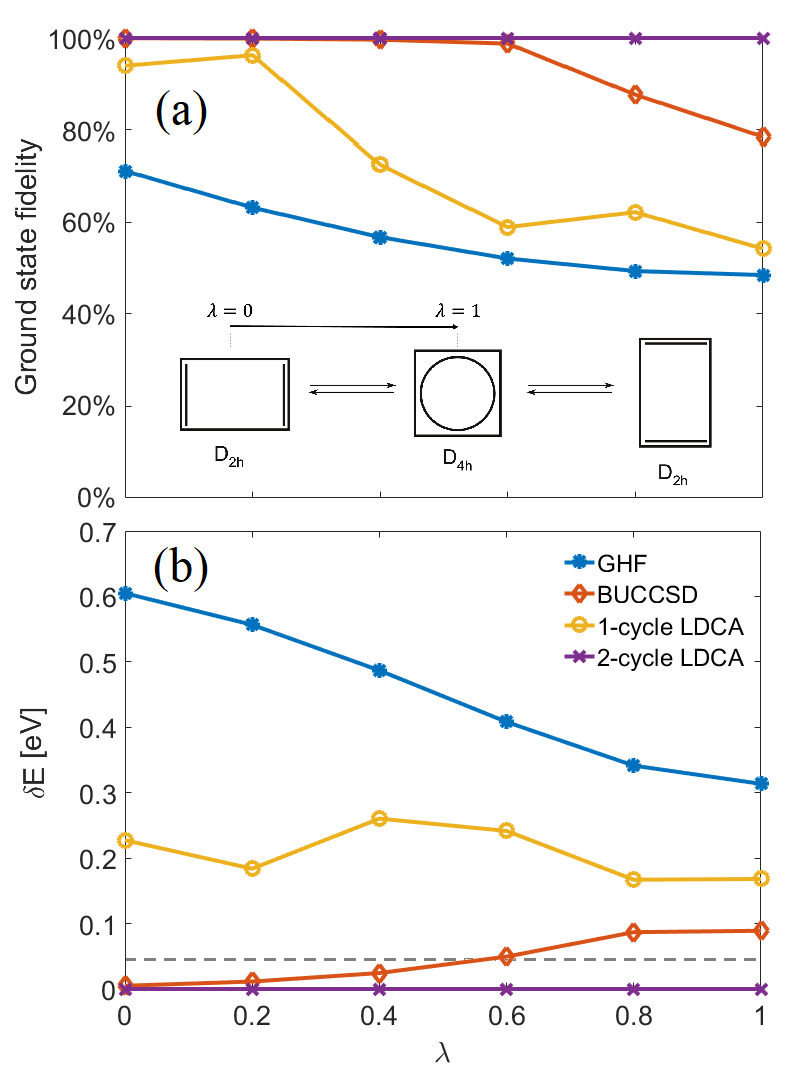}\caption{In (a), we show the fidelity of the ground state preparation along
the automerization reaction path of cyclobutadiene. Subfigure (b)
shows the difference with the exact ground state for the various ansatzes.
Chemical accuracy is approximately 0.043 eV (dashed line).\label{fig:Cyclobutadiene}}
\par\end{centering}
\end{figure}

Figure \ref{fig:Cyclobutadiene} compares the accuracies of different ansatzes for the cyclobutadiene automerization reaction. We observe that GHF ansatz is considerably
improved by BUCCSD close to the $D_{2h}$ ground state but the improvement is less prominent as we approach the strongly correlated $D_{4h}$ transition state. As in the $2\times2$
Fermi-Hubbard case, the 1-cycle LDCA method yields accuracies between those of GHF and BUCCSD while the 2-cycle LDCA method produces
the numerically exact ground state for all values of $\lambda$. This surprising result suggests that LCDA is potentially useful for treating cases of strong correlation in quantum chemistry.

\section{Discussion\label{sec:Discussion}}

The results presented in the previous section suggest that the LDCA could outperform other ansatzes employed for VQE calculations, such as BUCC, both in accuracy and efficiency. Being a method inspired by BUCC, the LDCA scheme inherits some properties of this ansatz. For instance, in the
limit of 1-cycle LDCA with all $\theta_{ij}^{ZZ\left(k,1\right)}$
set to zero, we recover the BUCC ansatz with single excitations. We point out that this choice of parameters cannot improve the GHF solution since it only amounts to a basis rotation
of the fermionic mode for which the Bogoliubov transformation has
already been optimized. Since the mapping \eqref{eq:NNRotations} between the Bogoliubov transformation and the matchgate circuit relies on the Jordan-Wigner transformation which associates Pauli strings of length $O\left(M\right)$ to fermionic operators, it may be possible to further reduce the length of the measured Pauli strings by working out a similar mapping in the Bravyi-Kitaev basis \cite{Bravyi02} where operators are represented by strings of length $O\left(\textrm{log} M\right)$. For completeness, we also numerically benchmarked the traditional
UCCSD scheme \citep{Peruzzo14,McClean16,Romero17} and found that it provides the same results as BUCCSD. This is expected
in the case of Hamiltonians with no explicit pairing terms. However,
such terms may appear in variational self-energy functional theory
\citep{Potthoff03,Potthoff06,Senechal08,Senechal08b,DDW16} where
fictitious pairing terms are added to a cluster Hamiltonian to recover
the magnetic and superconducting phase diagram in the thermodynamic
limit. 

Regarding the number of variational parameters, LDCA scales
as $O\left(LM^{2}\right)$ compared to $O\left(M^{4}\right)$ for UCCSD and BUCCSD with Gaussian basis set. There may exist constraints on the variational
parameters of LDCA that reduce their total number. To explore whether it was
possible to only measure $\left\langle H\right\rangle $ in the variational
procedure, we tried the ansatz with only number conserving terms (such
that all $\theta_{ij}^{XY\left(k,l\right)}=\theta_{ij}^{-YX\left(k,l\right)}=0$)
on the Fermi-Hubbard model but found a reduced overlap with the exact
ground state. This implies that a reconfiguration of the pairing amplitudes
with respect to the GHF reference state is an important condition
to reach an accurate ground state.

Our estimates of the circuit depth assume a quantum architecture consisting of a linear chain of qubits, which allows to maximize
the parallel application of gates through the algorithm. We leave
open the question of whether it is possible to achieve further improvements by using an architecture
with increased connectivity. We also assumed that nearest-neighbor
two-qubit gates could be implemented directly (as proposed for a linear chain of polar molecules \cite{Herrera14}). Although this is not the case on current ion trap and superconducting circuit technologies, the required gates can be implemented as long as tunable nearest-neighbor entangling
gates are available. In this case, only additional single-qubit basis rotation suffices, adding only a a small overhead in circuit depth \citep{Nielsen01}. 

Due to its better accuracy and reduced scaling in depth and number of parameters compared to previous ansatzes, we believe that the LDCA approach is a feasible alternative for studying strongly correlated systems in near-term quantum devices. In this case, we propose some strategies to ensure a better performance of the ansatz on real quantum processors with control
inaccuracies. For instance, we could calibrate the angles $\theta_{ij}^{\mu\nu}$
of the gate sequence of $U_{\mathrm{Bog}}^{\dagger}$ by minimizing the difference between the values of $\left\langle H\right\rangle $ and $\left\langle N\right\rangle$ measured on the quantum computer and the values obtained numerically for the GHF reference state. Similarly, it should be possible to experimentally estimate the errors on the energy and the number of particles for a given $L$-cycle LDCA by comparing the values of $\left\langle H\right\rangle $ and $\left\langle N\right\rangle$  obtained with all
$\theta_{ij}^{ZZ\left(k,1\right)}$ set to zero
with the exact classical results computed as described in section \ref{subsec:GHFT}. Instead of setting $\theta_{ij}^{ZZ\left(k,1\right)}$ to zero, one might also replace the $ZZ$
rotations with equivalent time delays.

Finally, we point out that our formalism should be general enough to implement the simulation
of nucleons \citep{Signoracci15}. However, we abstained from venturing in
the numerical simulation of such systems as it is beyond our fields
of expertise. Similarly, our method could be employed to study the ground state of gauge theories in the quantum link model \citep{Byrnes06,Zohar15,Dalmonte16}.

\section{Conclusion\label{sec:Conclusion}}

In this work, we generalized the Bogoliubov coupled cluster ansatz to a unitary
framework such that it can be implemented as a VQE scheme on a quantum
computer. We showed how the required GHF reference state can be computed
from the theory of fermionic Gaussian states. Those states include
Slater determinants used in quantum chemistry as well as mean field
superconducting BCS states. We described a procedure to prepare fermionic Gaussian states on a quantum computer using a circuit of nearest-neighbor matchgates with linear depth on the size of the system.
By augmenting the set of available gates with nearest-neighbor $\sigma_{z}\otimes\sigma_{z}$
rotations, we constructed a low-depth circuit ansatz (LDCA) that can systematically
improve the preparation of approximate ground states for fermionic
Hamiltonians. Each added cycle increases linearly the depth of the
quantum circuit, which makes it practical for implementations in near-term quantum devices.

We used a cluster of the
Fermi-Hubbard model and the automerization of Cyclobutadiene as examples to assess the accuracy of the BUCC and LDCA ansatzes. Our results showed that LDCA has the potential to accurately described the exact ground state of strongly correlated fermionic systems on a quantum
processor. In addition, our proposed BUCC and LDCA approaches can be used to approximate the ground states of Hamiltonians with pairing fields. This feature, not present in previous ansatzes such as UCC, extends the range of applicability of VQE to problems in condensed matter and nuclear
physics. Since the number of particles is not conserved in
BUCC and LDCA, we must impose constraints on the number of particles to carry out the optimization in the classical computer. Future work will be devoted to benchmarking the accuracy of the LDCA method for a larger variety of molecular systems and determining the scaling in the number of cycles required to describe the ground states of general systems.

\begin{acknowledgments}
We would like to thank Ryan Babbush for helpful discussions. Jonathan Romero and Al\'{a}n Aspuru-Guzik acknowledge the Air Force Office of Scientific Research for support under Award: FA9550-12-1-0046. Sukin Sim is supported by the DOE Computational Science Graduate Fellowship under grant number DE-FG02-97ER25308. Pierre-Luc Dallaire Demers and Al\'an Aspuru-Guzik acknowledge support from the Vannevar Bush Fellowship from the United States Department of Defense under award number N00014-16-1-2008 under the Office of Naval Research.
\end{acknowledgments}

\bibliographystyle{apsrev4-1}
\bibliography{Bogoliubov}

\end{document}